\documentclass[twocolumn]{cinc}

\usepackage{amsmath, graphicx, xurl, hyperref, amssymb, comment, booktabs, tabularx, multirow, rotating, adjustbox, bbold}
\usepackage[dvipsnames]{xcolor}
\newcolumntype{L}{>{\raggedright\arraybackslash}X}
\newcolumntype{C}{>{\centering\arraybackslash}X}

\begin{document}
\bibliographystyle{cinc}

\title{Estimation of f-wave Dominant Frequency Using a Voting Scheme}


\author {Shany Biton$^{1}$, Mahmoud Suleiman$^{2}$, Noam Ben Moshe$^{1,3}$, Leif Sörnmo$^{4}$, Joachim A. Behar$^{1}$ \\
\ \\ 
$^1$Faculty of Biomedical Engineering, Technion-IIT, Haifa, Israel \\
$^2$Department of Cardiology, Rambam Medical Center and Technion The Ruth and Bruce Rappaport Faculty of Medicine, Haifa, Israel\\
$^3$Faculty of Computer Science, Technion-IIT, Haifa, Israel\\
$^4$Department of Biomedical Engineering, Lund University, Lund, Sweden}

\maketitle
\begin{abstract}

\textbf{Introduction}: Atrial fibrillation (AF) is the most common heart arrhythmia, characterized by the presence of fibrillatory waves (f-waves) in the ECG. We introduce a voting scheme to estimate the dominant atrial frequency (DAF) of f-waves. \textbf{Methods}: We analysed a subset of Holter recordings obtained from the University of Virginia AF Database. 100 Holter recordings with manually annotated AF events, resulting in a total 363 AF events lasting more than 1~min. The f-waves were extracted using four different template subtraction (TS) algorithms and the DAF was estimated from the first 1-min window of each AF event. A random forest classifier was used. We hypothesized that better extraction of the f-wave meant better AF/non-AF classification using the DAF as the single input feature of the RF model.  \textbf{Results}: Performance on the test set, expressed in terms of AF/non-AF classification, was the best when the DAF was computed computed the three best-performing extraction methods. Using these three algorithms in a voting scheme, the classifier obtained AUC=0.60 and the DAFs were mostly spread around 6~Hz, 5.66 (4.83-7.47). \textbf{Conclusions}: This study has two novel contributions: (1) a method for assessing the performance of f-wave extraction algorithms, and (2) a voting scheme for improved DAF estimation.

\end{abstract}

\section{Introduction}

Atrial fibrillation (AF) is the most common arrhythmia with a prevalence that increases with age~\cite{chugh2014worldwide}. AF diagnosis using the surface ECG is mainly based on whether the RR interval pattern is irregular, P-wave are absent, and fibrillatory waves (f-waves) present, where RR interval irregularity stands out as the most prominent. However, other arrhythmias are also manifested by irregular RR interval patterns which resemble AF. Therefore, f-wave information is important to consider when distinguishing AF from other arrhythmias. However, f-wave extraction is complex since the atrial and ventricular activities overlap in both the spectral and the temporal domain. Several methods have been proposed to address this problem. When the number of leads is limited such as in Holter systems, average beat subtraction, also known as template subtraction (TS), is a widely used technique \cite{SornmoAFbookExtraction}. A disadvantage with this approach is its inability to handle non-stationary situations, leading to that the extracted f-wave signal is distorted and contains QRST residuals. 

The performance of f-wave extraction algorithms has been benchmarked by means of simulated ECG signals using local performance measures, including the root mean square error between the ground truth and the extracted f-wave signal~\cite{petrenas2012cancellation}. Other types of local performance measures quantify the amount of QRST-related residuals and the spectral concentration~\cite{sornmo2018atrial}. Another approach to assessing performance is to employ global measures, pursued in the present study together with the hypothesis that better f-wave extraction means better AF/non-AF classification. Classification is performed using the dominant atrial frequency (DAF) as the single input feature.

\section{Methods}
\subsection{Dataset}
The University of Virginia Atrial Fibrillation Database (UVAFDB) \cite{carrara2015heart, moss2014local} consists of 3-lead ECG Holter recordings acquired during the period 2004--2010 by the University of Virginia health system. Recordings were digitized at 200 Hz. The database contains 2,891 recording of patients, with rhythm annotation automatically generated by the Philips Holter software. In the present study, we use a subset of 100 recordings from patients older than 18 years, each recording manually reviewed and annotated with respect to AF events by an expert cardiologist as described in~\cite{Biton2022gengeralization}. The median and interquartile patient age were 69.0 (58.8-76.2), each was 24h long. A total of 457 non-overlapping AF events lasting 128.0 (68--718) seconds was analyzed.

\subsection{Preprocessing}
Similar to the work in~\cite{chocron2020remote}, we computed the signal quality index bSQI \cite{behar2013ecg, li2007robust, gendelman2021physiozoo} and discarded segments when below~0.8. ECG recordings were prefiltered using a zero-phase second-order infinite impulse response bandpass filter, with a passband of 0.67--100~Hz to remove baseline wander and high-frequency noise~\cite{kligfield2007recommendations}. A notch filter at 60~Hz was used to remove power-line interference. Channel 1, approximating a modified lead V1 was used as it is the most informative lead for expert annotation.

\subsection{F-wave extraction}
In this work, we evaluate four TS techniques for single-lead f-wave extraction. The algorithms were originally implemented in the context of fetal ECG extraction from the abdominal fetal-maternal mixture \cite{behar2014combining}. These algorithms are similar to those used in the context of f-wave extraction as they aim to cancel out the pseudo-periodic components in phase with the (maternal) cardiac cycle, while leaving the fetal ECG (in case of fetal ECG) or the f-waves (in case of AF) in the residual signal. 

The first algorithm is the basic TS adapted from \cite{slocum1992diagnosis} denoted as T$_{B}$: A QRST template is obtained by averaging temporal QRST windows with respect to the duration of the whole cardiac cycle. The QRST template is then subtracted from the original ECG signal, resulting in a residual signal that contains the f-wave and residual noise. The other three TS variants offer different degrees of adaptability of the template: in TS$_{CE}$ the QRST template is scaled by a gain before subtraction~\cite{beckers2005determination}; in TS$_{SU}$ the scaling procedure is performed for the P-wave, the QRS-complex and the T-wave individually~\cite{martens2007robust}; in TS$_{PCA}$ (principal component analysis, PCA) the almost periodic characteristics of the ECG is used for selective separation of the QRST and the f-waves~\cite{kanjilal1997fetal}.

\begin{figure*}[h]
\centering
\includegraphics[page=1,width=1\textwidth]{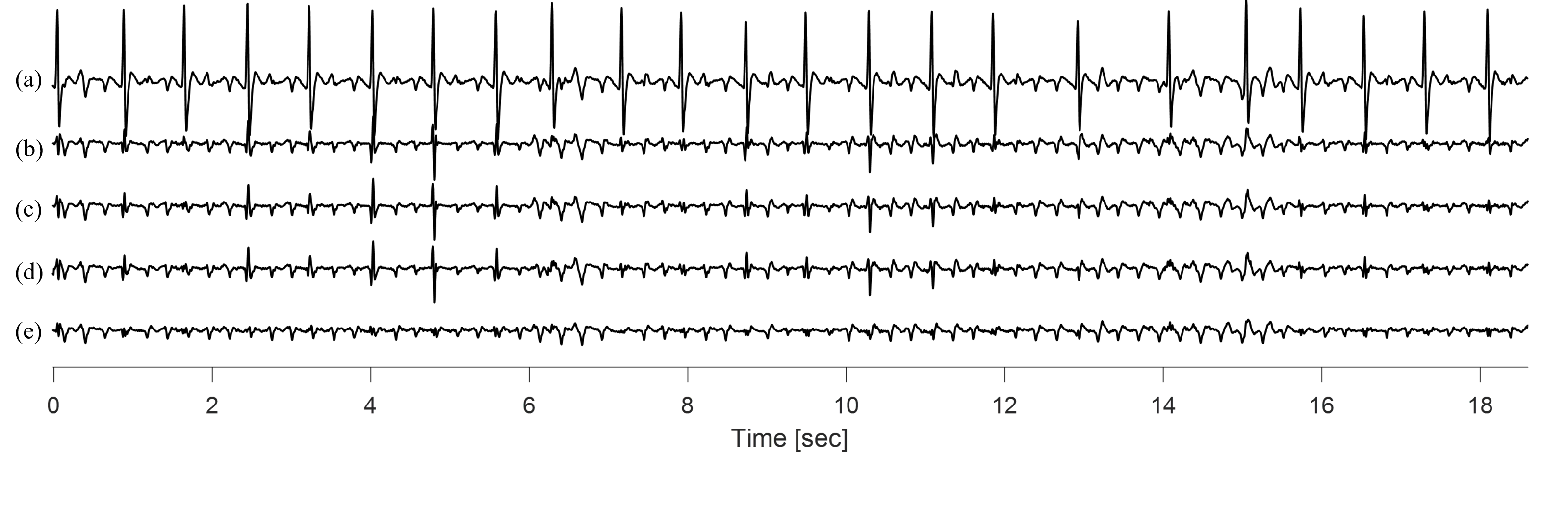}
\caption{Example of f-wave extraction. (a) A single-lead ECG in AF and related extracted f-waves obtained using (b)~TS$_{B}$, (c)~TS$_{CE}$, (d)~TS$_{SU}$ and (e)~TS$_{PCA}$.}
\label{exampleFwave}
\end{figure*}

The algorithms are listed in Table \ref{tab:ExtractionAlgorithms}. Following the preprocessing step, AF events were extracted, then f-wave extraction algorithms were run only on the first minute of each event. We used 1-min windows similar to the work in~\cite{alcaraz2009non}. Matlab R2021a and Python 3.8 were used. An example of the extracted f-wave is shown in Fig~\ref{exampleFwave}.

\subsection{Machine learning}
A random forest (RF) classifier was used. We extracted similar number of non-AF and AF windows for which the DAF was computed. A train-test split of 80\%--20\% was performed with equal number of AF/non-AF windows in the training set. To asses the performance of the trained RF in classifying AF and non-AF windows correctly, the following statistics were computed: the harmonic mean between the sensitivity and positive predictive value, denoted F$_{1}$, and the area under the receiver operating characteristic (AUROC). The default decision threshold was used for classification in training and testing.

\begin{table}[htbp]
\label{tab:ExtractionAlgorithms}
\caption{ Summary of the f-wave extraction algorithms experimented in this research. They were implemented in the previous work from Behar, J. et al. \protect\cite{behar2013ecg}.}
\vspace{4 mm}
\centerline{\begin{tabular}{lcr} \hline\hline
Method & Reference \\ \hline
TS$_{B}$   &  \cite{cerutti1986variability} \\
TS$_{CE}$  & \cite{cerutti1986variability}\\
TS$_{SU}$  & \cite{martens2007robust}\\
TS$_{PCA}$ & \cite{kanjilal1997fetal} \\
\hline\hline
\end{tabular}}

\end{table}

\section{Frequency estimation}
\begin{figure}[h]
\includegraphics[page=1,width=0.44\textwidth]{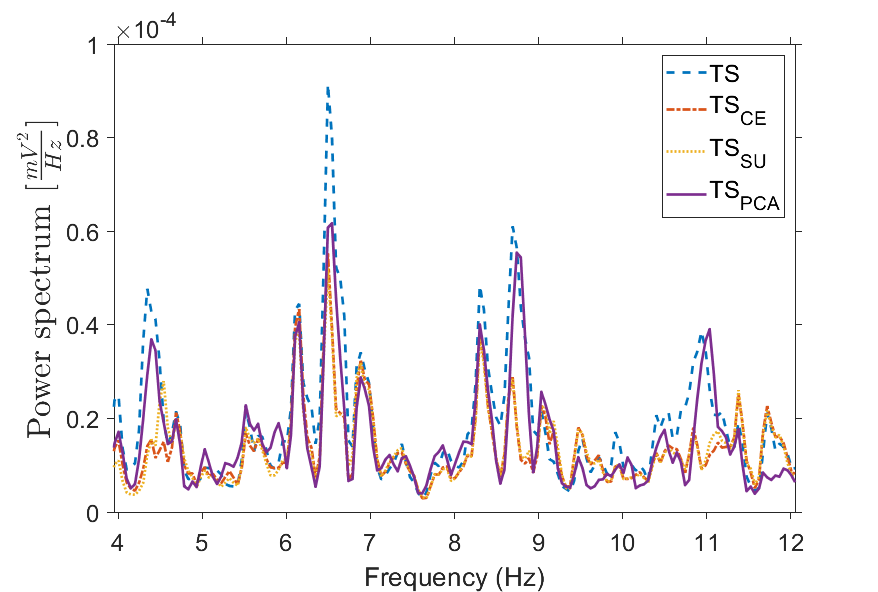}
\caption{Example of power spectra for extracted f-wave signals obtained by the four different algorithms.}
\label{examplePS}
\end{figure}

Spectral analysis of the extracted f-wave signals was applied to estimate the DAF, defined as the position of the largest peak in the interval of 4--12~Hz of the power spectrum of the extracted f-wave signal~\cite{SornmoAFbookExtraction}. The power spectrum was computed using Welch's method with a Hamming window and 50\% segment overlap. An example of power spectrum is shown in Fig.~\ref{examplePS}. After extracting the DAF using the different algorithms, a voting scheme was applied to determine the final DAF, i.e., the median value was taken as the DAF across the best performing methods for AF classification. Since real data were analyzed, a comparison of the resulting DAF to that of a reference f-wave signal could not be done.

\section{Results} 
No recordings were excluded from the dataset due to low quality. 476 AF event were extracted with length of 30 seconds defined to be the minimal length of AF episode as in \cite{chocron2020remote}. Out of those,  94 (20.57\%) events shorter then 1-min were excluded, as windows of 1-min length were used. Results are shown for total of 363 AF windows belonging to 72 recordings. 

\begin{figure}[h]
\includegraphics[page=1,width=1\linewidth]{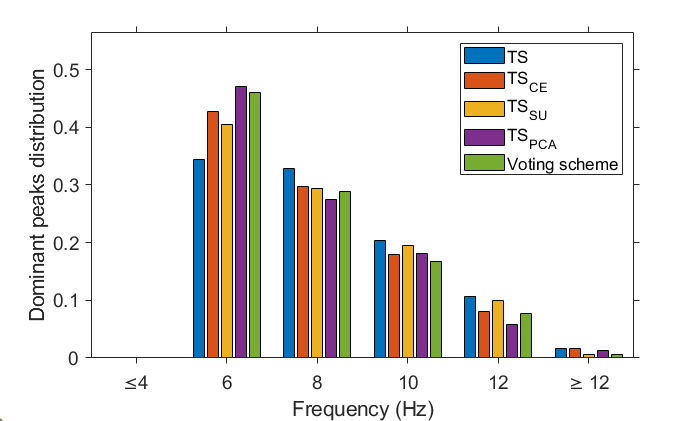}
\caption{Distribution of the DAF obtained using four different f-wave extraction algorithms for AF examples.}
\label{DAFperAlg}
\end{figure}

The DAFs were mostly spread within the interval 6--8~Hz: TS$_{B}$ 6.1 (5.04--8.59), TS$_{CE}$ 5.81 (4.88--7.91), TS$_{SU}$ 5.96 (4.88--8.20) and TS$_{PCA}$ 5.62 (4.83--7.56). See the distributions in Fig.~\ref{DAFperAlg}.

The test set classification performances for the DAF estimation using TS$_{B}$, TS$_{SU}$ or TS$_{CE}$ were AUROC of 0.59, 0.59 and 0.59, respectively (Table~\ref{tab:RFscores}). The voting scheme involving the three best performing techniques obtained an F$_{1}$ of 0.63 and an AUROC equal to~0.60. Using the voting scheme, the DAF was mostly spread around median, (Q1-Q3) 6~Hz (4.87--7.47).

\begin{table}[htbp]
\caption{\label{tab:RFscores} Performance statistics for different RF models.}
\vspace{4 mm}
\centerline{\begin{tabular}{lcc} \hline\hline
Method     & F$_{1}$ & AUROC \\ \hline
TS$_{B}$   & 0.62       & 0.59 \\
TS$_{CE}$  & 0.61       & 0.59 \\
TS$_{SU}$  & 0.61       & 0.59 \\
TS$_{PCA}$ & 0.56       & 0.53 \\
Voting scheme & \textbf{0.63} & \textbf{0.60} \\

\hline\hline
\end{tabular}}

\end{table}

\section{Discussion}  
The main contribution of this research is the introduction of a novel method for assessing f-wave extraction performance, whereas, in previous research, the emphasis has been on assessing local performance of f-wave extraction. An interesting approach was to use simulated ECG signals as it provides access to the ground truth, however, simulation has the disadvantage of not fully account for the physiological variability across patients. As a consequence, we seek to develop a new method for assessing the performance of f-wave extraction algorithms based on real data experiments where the ground truth DAF is unknown. 

Another important contribution of this study is the introduction of a voting scheme for DAF estimation based on the three best performing algorithms, i.e., TS$_{B}$, TS$_{SU}$ and TS$_{CE}$. 

This work has important limitations. First, the dataset is limited to 100 patients all from a single cardiac center. Future work needs to evaluate the performance of the algorithms on datasets from multiple centers, across ethnicity, age and gender. A second limitation is that we focused on the extraction of the DAF as the sole feature for assessing the performance of f-wave extraction. The usage of this single feature may not be enough to best rank the f-wave extraction algorithms using our data-driven method. Other f-wave features such as amplitude may also be taken into account to supplement the DAF in training the RF classifier. In that respect the ranking of the algorithms could change. Finally, other extraction algorithms could be included~\cite{SornmoAFbookExtraction}, e.g., noise-dependent weighted averaging, adaptive filtering and Kalman filter based methods~\cite{behar2014combining}.

\balance

\section*{Acknowledgments}  
This research was supported for SB, NBM and JAB by a grant (3-17550) from the Ministry of Science \& Technology, Israel \& Ministry of Europe and Foreign Affairs (MEAE) and the Ministry of Higher Education, Research and Innovation (MESRI) of France. SB, NBM, MS and JAB acknowledge the support of the Technion-Rambam Initiative in Artificial Intelligence in Medicine.

\bibliography{refs}

\begin{correspondence}
Joachim A. Behar\\
jbehar@technion.ac.il
\end{correspondence}

\end{document}